# Tailored plasmons in pentacene/graphene heterostructures with interlayer electron transfer

F. Hu[1,2]*, M. Kim[1,2]*, Y. Zhang[3], Y. Luan[1,2], K. M. Ho[1,2], Y. Shi[3], C. Z. Wang[1,2]†, X. Wang[3]†, Z. Fei[1,2]†

[1]Department of Physics and Astronomy, Iowa State University, Ames, Iowa 50011, USA
[2]U.S. DOE Ames Laboratory, Iowa State University, Ames, Iowa 50011, USA
[3]National Laboratory of Solid State Microstructures, School of Electronic Science and Engineering and Collaborative Innovation Center of Advanced Microstructures, Nanjing University, Nanjing 210093, China

* These authors contributed equally to this work.

† C.Z.W. (wangcz@ameslab.gov); X.W. (xrwang@nju.edu.cn); Z.F. (zfei@iastate.edu)

**Abstract**

Van der Waals (vdW) heterostructures, which are produced by the precise assemblies of varieties of two-dimensional (2D) materials, have demonstrated many novel properties and functionalities. Here we report a nano-plasmonic study of vdW heterostructures that were produced by depositing ordered molecular layers of pentacene on top of graphene. We find through nano-infrared (IR) imaging that surface plasmons formed due to the collective oscillations of Dirac fermions in graphene are highly sensitive to the adjacent pentacene layers. In particular, the plasmon wavelength declines systematically but nonlinearly with increasing pentacene thickness. Further analysis and density functional theory (DFT) calculations indicate that the observed peculiar thickness dependence is mainly due to the tunneling-type electron transfer from pentacene to graphene. Our work unveils a new method for tailoring graphene plasmons and deepens our understanding of the intriguing nano-optical phenomena due to interlayer couplings in novel vdW heterostructures.

**Main text**

Graphene plasmons are collective oscillations of Dirac quasiparticles in graphene with many desirable characteristics including high spatial confinement, long lifetime, broad spectral range, and electrical tunability.[1-19] These unique properties make graphene a good candidate for varieties of plasmonic applications that are not accessible by conventional plasmonics based on noble metals. Despite the above merits, the plasmonic properties and functionalities of graphene alone are still limited. One convenient way to engineer graphene plasmons is by constructing van der Waals (vdW) heterostructures using atomic layers of graphene and other two-dimensional (2D) materials. Indeed, the 2D nature of graphene makes it extremely sensitive to interlayer couplings that could modify dramatically the properties of plasmons. Recent studies have explored a variety of new plasmonic phenomena in graphene-based vdW materials and heterostructures, where the coupling mechanisms are mainly plasmon-phonon interactions[20-23] and moiré superlattice modulations[24,25].

Here we report interlayer electron transfer as a new mechanism that can be used to tailor surface plasmons in graphene. The samples studied here are pentacene/graphene vdW heterostructures prepared by physical vapor transport deposition of uniform pentacene molecular layers on graphene, and they are sitting on the standard $SiO_2$/Si substrates. Detailed introductions about the sample growth and device fabrication procedures are introduced in the previous study.[26] Note that molecule/graphene vdW heterostructures have been widely studied in recent years and have demonstrated many superior electronic and optoelectronic properties.[26-28] In this work, we report a comprehensive experimental and theoretical study of the plasmonic responses of the pentacene/graphene heterostructures.

To perform nano-IR studies of the plasmonic responses of the heterostructure samples, we utilized the scattering-type scanning near-field optical microscope (s-SNOM) that is built on a tapping-mode atomic force microscope (AFM). As illustrated in Figure 1a, the sharp metalized AFM tip is illuminated by a *p*-polarized IR laser beam, thus generating an intense electric field underneath the tip apex due to the so-called 'lightening-rod effect'. Such a strong field is highly confined in space and possesses a wide range of in-plane momenta ($q$), which enables direct optical excitation and detection of graphene plasmons. The IR detector collects scattered photon signals off the coupled tip-sample system. Demodulating the signals at high harmonics of the AFM tapping frequency strongly suppresses the background signal. Furthermore, we implement a pseudo-heterodyne interferometer that allows us to extract both the amplitude and phase components of the IR signal. In the current work, we discuss mainly the IR amplitude ($s$) signal that scales monotonically with the plasmon field amplitude right underneath the tip.[29] All our experiments were performed at ambient conditions.

Figure 1b plots the AFM topography image of a typical pentacene/graphene heterostructure sample, where clear staircase features are seen in the field of view, corresponding to pentacene layers with different thicknesses. By measuring the thickness of different regions of the sample (Figure 1d) based on the AFM line profiles (Figure S1), we can accurately determine the number of pentacene layers as labeled in Figure 1b,c. According to the previous study[26], the orientation of pentacene molecules shows variations from layer to layer close to the graphene interface due to the competition between the molecule-graphene interactions and the intermolecular interactions. More specifically, as shown in Figure 1a, pentacene molecules first form a sheet of the flat-lying wetting layer (WL) on graphene, then the inclined one layer (1L) following by the upright-standing two layers (2L) and few layers (3L, 4L, etc.). The orientation angles of pentacene molecules to the substrate are 0°, 61° and 82° for the flat-lying, inclined and standing layers, respectively. As discussed in detail below, the molecule orientation plays a critical role in the overall plasmonic responses of pentacene/graphene heterostructures. Note that bare WL pentacene could exist in freshly grown samples, but it will soon disappear after exposure to air due to the dewetting, suggesting that upright-standing molecules are possibly more favorable energetically in air than the flat-lying molecules[30]. Thicker pentacene layers with standing molecules are generally more stable and can last for many days at ambient conditions thus suitable for systematic nano-IR studies.

In Figure 1c, we present the s-SNOM imaging data of the sample shown in Figure 1b, where we plot the IR amplitude normalized to that of the SiO$_2$/Si substrate. The laser energy is set to be $E$ = 116 meV that is away from the strong optical phonon resonance of SiO$_2$ (~140 meV), so the IR responses at this energy are predominantly due to graphene plasmons. As shown in Figure 1c, there is a clear IR signal contrast between different pentacene layers on graphene. For quantitative analysis, we plot in Figure 1e the IR amplitude versus the number of pentacene layers, which indicates a systematic decrease of the IR amplitude with increasing pentacene thickness. Moreover, we found a relatively big drop of IR amplitude signal from 1L to 2L pentacene (~27%), but only slight declines from 2L to 3L pentacene (~9%) and from 3L to 4L pentacene (~4%). Similar pentacene thickness dependent signal variation is also seen in other samples (e.g. sample 2 in Figure S4).

From Figure 1c, we also observed a bright edge feature surrounding the sample. To reveal the details about the bright edge feature, we performed high-resolution s-SNOM imaging measurements close to the sample edge (Figure 2a-e), where we observed bright fringe(s) parallel to the sample edge. According to previous studies[15,16], these bright fringes are generated due to the constructive interference between tip-launched and edge-reflected surface plasmons of graphene. The plasmonic origin of these fringes is verified by frequency-dependence studies (Figures S2 and S3). In addition to the bright fringes, we also occasionally see weak oscillations of signals distributed along the sample edge, for example in Figure 2c,d. These edge oscillations are generated due to scattering and interference of one-dimensional edge plasmons and they normally appear at relatively rough edges (e.g. in the case of Figure 2c,d) or close to sharp corners.[31,32]

Now we wish to perform quantitative analysis on the imaged plasmon fringes. For that purpose, we plot in Figure 2f-j the line profiles (grey curves) extracted perpendicular to the fringes in Figure 2a-e, respectively. From both the IR amplitude images and profiles, we found a systematic variation of the plasmon fringes with pentacene thickness. First, the samples with thicker layers of pentacene show weaker fringe intensity, and the strongest fringe is observed in bare graphene. In addition, the width of the bright fringe decreases with increasing pentacene thickness, implying a reduction of plasmon wavelength. Furthermore, the number of fringes decreases with increasing pentacene thickness. For example, there are at least 3 bright fringes at the edge of bare graphene, 2 clear bright fringes in the case of 1L pentacene on graphene, and only 1 clear fringe for 2L, 3L and 4L pentacene on graphene. The decrease of the fringe number indicates an increase in the plasmon damping rate.

The fringe profiles shown in Figure 2f-j allow us to fit quantitatively the complex plasmon wavevector $q_p = q_1 + iq_2$ of graphene, based on which we can determine the plasmon wavelength ($\lambda_p = 2\pi/q_1$) and damping rate ($\gamma_p = q_2/q_1$). To perform the fit, we adopted a quantitative s-SNOM model that approximates the s-SNOM tip as a conducting spheroid (Figure 3a). This model calculates accurately the s-SNOM signals by evaluating the total radiating dipoles ($p_z$) of the tip-sample system. By computing $p_z$ at multiple $x$ and $z$ coordinates of the tip, we were able to obtain line profiles of s-SNOM signals with quantitative accuracy. More introductions about the model are given in the Supporting Information. The same model has been applied to calculate the plasmon fringe profiles of bare graphene and other graphene-based vdW materials and heterostructures reported in earlier works.[15, 33-35]

The modeling profiles are plotted in Figure 2f-j as red dashed curves, which show good consistency with the experimental data profiles (grey). The $\lambda_p$ and $\gamma_p$ parameters determined through the fitting are given in Figure 3b,c, respectively. Figure 3b indicates that $\lambda_p$ decreases systematically with increasing pentacene thickness. For example, from bare graphene to 4L pentacene on graphene, $\lambda_p$ drops from 250 nm to 205 nm. More interestingly, $\lambda_p$ shows a sharp reduction from 240 nm to 215 nm when the pentacene thickness changes from 1 layer to 2 layers. This sharp reduction of $\lambda_p$ also results in an abrupt drop of the overall IR amplitude signal from 1L to 2L pentacene (Figures 1e and Figure S4). The plasmon damping rate $\gamma_p$, on the other hand, increases systematically with pentacene thickness, which is consistent with the decrease of the number of plasmon fringes shown in Figure 2. As shown in Figure 3c, the extracted $\gamma_p$ by fitting the plasmon fringe profiles increases from 0.14 for bare graphene to 0.17, 0.24, 0.27 and 0.3 when adding 1L, 2L, 3L and 4L pentacene on top of graphene, respectively. Like $\lambda_p$, $\gamma_p$ also undergoes a larger change from 1L to 2L pentacene (~0.07) compared to that between other adjacent layers (~0.03).

We now elaborate on the possible causes of the observed thickness dependence of the plasmonic parameters, Under the Drude and long-wavelength approximations, the plasmon wavevector $q_p$ can be written as [15,16,25]

$$q_p \equiv q_1 + iq_2 \approx \frac{2\pi\varepsilon_0\kappa}{e^2 E_F} E(E + iE_\Gamma), \tag{1}$$

where $e$ is the elementary charge, $E_F$ is the Fermi energy of graphene, $E_\Gamma$ is scattering energy of Dirac Fermions in graphene, and $\kappa = \kappa_1 + i\kappa_2$ is the effective dielectric constant of the environment of graphene ($\kappa_1$ and $\kappa_2$ are the real and imaginary parts of $\kappa$). For bare graphene, $\kappa$ is an average value from the dielectric constants of air and SiO$_2$: $\kappa = (1 + \varepsilon_s)/2$ ($\varepsilon_s \approx 4.4 + 0.3i$ at $E = 116$ meV). In the case of pentacene/graphene heterostructures, dielectric constants of pentacene also contribute to $\kappa$. Note that our graphene samples are highly doped at ambient conditions with $E_F$ above 0.4 eV (see discussions below), so contributions from interband transitions at our energy regime are negligible thus not considered here. From eq 1, one can obtain the plasmon wavelength $\lambda_p = 2\pi/q_1$:

$$\lambda_p = 2\pi / q_1 \approx e^2 E_F / (\varepsilon_0 \kappa_1 E^2). \tag{2}$$

Therefore, the observed layer dependence of $\lambda_p$ (Figure 3b) is possibly due to the change of $E_F$ of graphene and/or the dielectric constants of pentacene. Note that eqs 1 and 2 are mainly for discussions of general physics of graphene plasmons. We used the transfer matrix method to compute numerically the plasmon dispersion and plasmon wavelength of the entire pentacene/graphene/substrate system (Supporting Information).

We first evaluate the effects solely due to the dielectric screening of pentacene layers with a fixed $E_F$ of 0.47 eV — the Fermi energy of bare graphene accurately determined by fringe profile fitting (Figure 2f). The large $E_F$ indicates the high hole doping of graphene on SiO$_2$, which is originated from the vacuum annealing during the pentacene growth process followed by days of air exposure.[36,37] The anisotropic dielectric constants of pentacene with different thicknesses were calculated from density functional theory (DFT) calculations (Supporting Information), which vary from 2.1 to 2.7 in the *ab* plane and from 1.3 to 2.6 along the *c*-axis for different pentacene thicknesses. Note that our excitation laser energy (116 meV) is away from the strong vibrational resonances of pentacene (the nearest strong resonance is at 112 meV with a resonance width of about 0.4 meV)[38,39], so the vibrational modes of pentacene do not affect graphene plasmons. The calculated $\lambda_p$ of graphene with a fixed $E_F$ under various pentacene layers is plotted in Figure 3b as blue triangles, which show a gentle and systematic decline with layer number ($\Delta\lambda_p \approx$ -4 nm on average per layer). Based on Figure 3b, we know that dielectric screening of pentacene alone cannot explain the sharp drop of $\lambda_p$ as pentacene thickness increases from 1 to 2 layers. The inconsistency between experimental and calculated $\lambda_p$ assuming a fixed $E_F$ indicates that the layer dependence of doping must be taken into consideration. Indeed, layer-dependent $E_F$ can be obtained accurately by fitting the experimental $\lambda_p$. As shown in Figure 3d, graphene under 1L pentacene has slightly smaller $E_F$ (~ 0.46 eV) compared to that of bare graphene (~ 0.47 eV). The $E_F$ of graphene under 2 to 4 layers pentacene is much lower, down to ~ 0.42 eV.

The unique pentacene layer dependence of $E_F$ (Figure 3d) is, in fact, originated from the charge transfer between graphene and pentacene. Charge transfer phenomena have also been observed at the interfaces between graphene and other types of molecules (e.g. C60, CNT, etc.).[40-42] To understand the transfer process here, we plot in Figure 4 the energy alignment diagrams between the graphene Fermi level (dashed line) and the highest occupied molecular orbits (HOMO) level of pentacene layers. The lowest unoccupied molecule orbits (LUMO) are ~2 eV above the HOMO level (not shown in Figure 4), so there are no unoccupied states available in pentacene close to the Fermi level of graphene. In Figure 4, we label the ionization potential (IP) values of graphene and pentacene, which is the energy difference from the Fermi level of graphene or HOMO energy of pentacene to the vacuum energy. Considering that graphene on $SiO_2$ is hole doped at ambient conditions[36,37], the IP can be calculated to be around 5.03 eV by adding the Fermi energy (~0.47 eV, Fermi level to Dirac point) and the work function of neutral graphene (~4.56 eV, Dirac point to vacuum energy).[42,43]

The IP of pentacene layers is sensitively dependent on the molecule orientation.[44-46] To obtain the IP values of pentacene layers, we performed first-principles electronic structure calculations based on DFT using the Vienna ab initio simulation package.[47,48] The atomic structures of the pentacene layers (Figure S6) are adopted from the previous study.[26] Such DFT calculations tend to underestimate the value of IP[49], so we considered the GW correction ($\Delta_{GW}$). Detailed introductions about the IP calculations are given in the Supporting Information. The final IP values of WL, 1L, 2L, and 3L pentacene without and with GW corrections are summarized in Table 1, where one can see a big drop (~0.77 eV) of IP from WL to 1L pentacene, followed by a small drop (~0.18 eV) from 1L to 2L pentacene. Starting from 2L pentacene and above, IP stays constant at 4.78 eV. Such a layer dependence is originated from the difference of the orientation angles of pentacene molecules (0°, 61° and 82° for the WL, 1L, and 2L or above, respectively). Our IP calculations are consistent with previous experimental results.[45,46] Note that the interface dipole between graphene and pentacene induces a shift of pentacene vacuum level by $\Delta \approx$ 0.1 eV.[46]

| Pentacene layers | $IP_{DFT}$ (eV) | $IP_{DFT} + \Delta_{GW}$ (eV) |
|---|---|---|
| WL | 4.76 | 5.72 |
| 1L | 3.99 | 4.95 |
| 2L | 3.81 | 4.77 |
| 3L | 3.82 | 4.78 |

**Table 1**. The calculated ionization potential (IP) of pentacene with different thicknesses.

Based on Figure 4, one can see that the Fermi level of graphene is much higher than the HOMO energy level of WL pentacene, so charge transfer between graphene and WL pentacene is forbidden. For 1L to 4L pentacene, HOMO energy level rises above $E_F$ of graphene, so electron transfer from pentacene to graphene is enabled. The amount of electron transfer from 1L pentacene to graphene is much less compared to that from thicker pentacene layers. For 1L pentacene, the reduction of graphene $E_F$ due to the charge transfer ($\Delta E_F$) is about 0.013 eV, corresponding to the change of carrier density ($\Delta n$) of about $0.9 \times 10^{12}$ cm$^{-2}$. For 2L to 4L pentacene, the resulting $\Delta E_F$ is about 0.047 eV, corresponding to the $\Delta n \approx 3.1 \times 10^{12}$ cm$^{-2}$. The size of $\Delta E_F$ is mainly due to the potential difference between graphene and pentacene ($\Delta E_{IP}$) (Figure 4). The amount of $\Delta E_{IP}$ for 1L pentacene (~0.08 eV) is much smaller than those of 2L to 4L pentacene (~0.25 eV). Another relevant factor is the density of states of pentacene layers. In principle, few-layer pentacene should have more electrons to offer compared to 1L pentacene. Note that the electron transfer discussed here is a tunneling process due to the presence of the WL pentacene that acts as a tunneling barrier (Figure 4) with a thickness of about 0.5 nm.[26] Effects of electron tunneling on surface plasmons have been studied in metal-molecule junctions, where unique quantum plasmonic responses were observed.[50] It is also proposed that electron tunneling can be utilized to generate graphene plasmons.[51,52] Therefore, the molecule/graphene heterostructure with interlayer electron tunneling studied here provides a unique platform to explore further the role of electron tunneling on graphene plasmons.

Finally, we wish to discuss the dependence of plasmon damping rate $\gamma_p$ on pentacene thickness. As discussed above (Figure 3c), $\gamma_p$ increases with pentacene thickness, and the increment of $\gamma_p$ is larger from 1L to 2L pentacene (~0.07) compared to that between other adjacent layers (~0.03), which implies a possible link between electron transfer and plasmon damping. Based on eq 1, we know that $\gamma_p$ can be written approximately as:

$$\gamma_p = q_2 / q_1 \approx \kappa_2 / \kappa_1 + E_\Gamma / E, \qquad (3)$$

which indicates that $\gamma_p$ originates from both the loss due to the dielectric environment and the scattering of Dirac fermions in graphene. As discussed above, $\kappa \approx 2.7 + 0.15i$ for graphene sitting directly on SiO$_2$ at $E = 116$ meV. As a semiconductor, pentacene behaves like a good dielectric with a negligible imaginary part of permittivity at the mid-IR region if it is away from the vibrational modes[39], so pentacene itself has little contribution to $\kappa_2$ at our excitation energy. Therefore, the enhanced $\gamma_p$ when adding pentacene layers is most likely due to the scattering of graphene carriers by impurities or localized charges in pentacene. With electron transfer, additional localized charges could be introduced to pentacene, which cause higher damping to graphene plasmons (Figure 3c). Increased charge scattering leading to a lower carrier mobility has been observed previously in transport studies of C$_{60}$/graphene heterostructures, where charge transfer was also involved.[53]

In summary, we have performed the first nanoplasmonic study of vdW heterostructures formed by organic 2D materials and graphene. By using the nano-IR imaging technique, we discovered that the graphene plasmons could be tailored by depositing molecule layers of pentacene on graphene. Unlike electrical gating that requires a constant bias voltage, the molecular deposition method is suitable for creating heterostructure samples or devices with tailored permanent properties for long-term applications. Through quantitative analysis and DFT calculations, we proved that the pentacene-layer dependence of graphene plasmons is mainly due to tunneling-type electron transfer from pentacene to graphene. Moreover, we found the electron transfer process is determined by the molecule orientation of each pentacene layer. Such a unique sensitivity to molecular orientations is highly desired for structural characterizations of molecules and bio-nanoparticles. Of course, the studies should not be limited to pentacene/graphene heterostructures. We expect more interesting nano-optical properties and functionalities to be discovered in heterostructures formed by graphene with other types of molecules. Our work broadens the understanding of the interlayer interactions of graphene with biomolecules and opens the door to future studies and applications of molecule/graphene heterostructures in nanophotonics and optoelectronics.

**Acknowledgments**
Work done at Ames Lab was supported by the U.S. Department of Energy, Office of Basic Energy Science, Division of Materials Sciences and Engineering. Ames Laboratory is operated for the U.S. Department of Energy by Iowa State University under Contract No. DE-AC02-07CH11358. The nano-optical imaging set-up was partially supported by the W. M. Keck Foundation. X.W. acknowledges the funding support from the National Natural Science Foundation of China 61734003 and the National Key Basic Research Program of China 2013CBA01604.

**Figures**

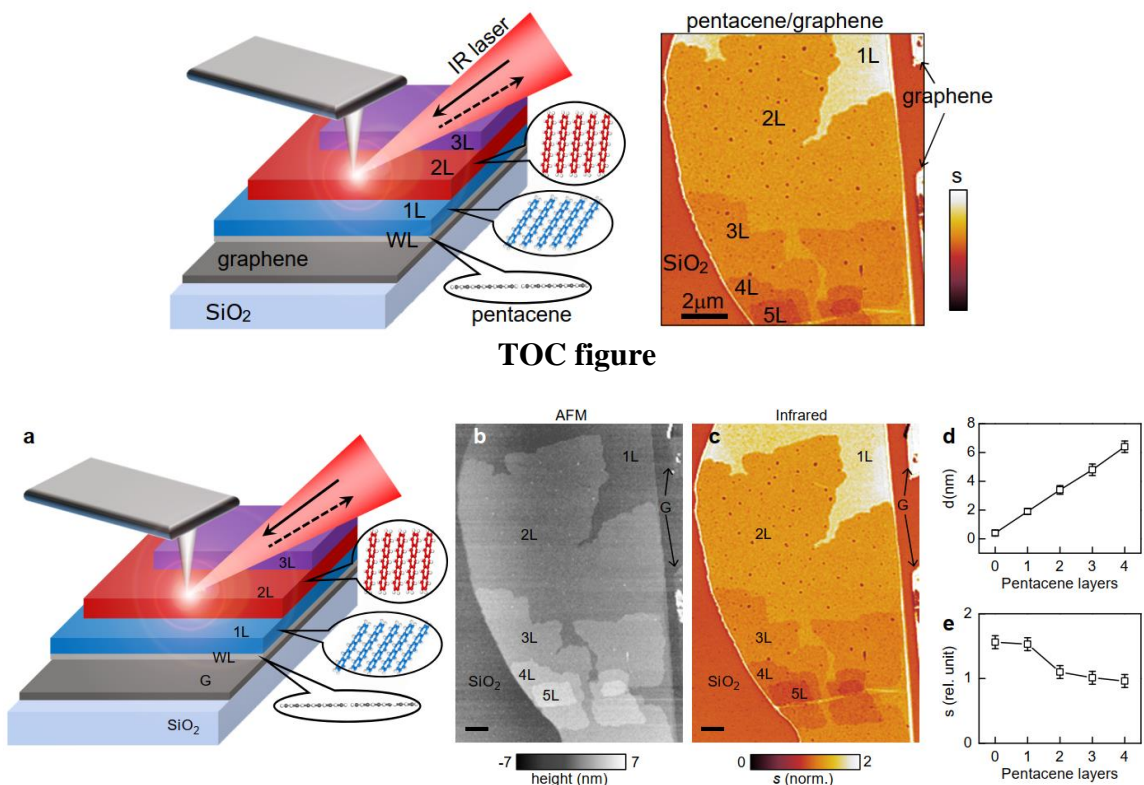

**TOC figure**

**Figure 1. a**, Illustration of the s-SNOM study of a pentacene/graphene heterostructure. The arrows sketch the incident laser and back-scattered photons, respectively. **b**, The AFM topography image of a typical pentacene/graphene heterostructure sample. **c**, The IR amplitude (*s*) image of the pentacene/graphene heterostructure taken at a photon energy of $E = 116$ meV. Here we normalized the amplitude signal to that of $SiO_2$. The labeling 'WL' represents the wetting layer on graphene, '1L'-'4L' represent 1-layer to 4-layer pentacene on graphene, and 'G' represents bare graphene. The scale bars: 1 μm. **d**, The thickness (d) of sample relative to the $SiO_2$ surface versus the number of pentacene layers. **e**, The IR amplitude signal versus the number of pentacene layers measured at locations away from the edge of the sample in Figure 1c. In **d** and **e**, the 0 pentacene layer corresponds to bare graphene.

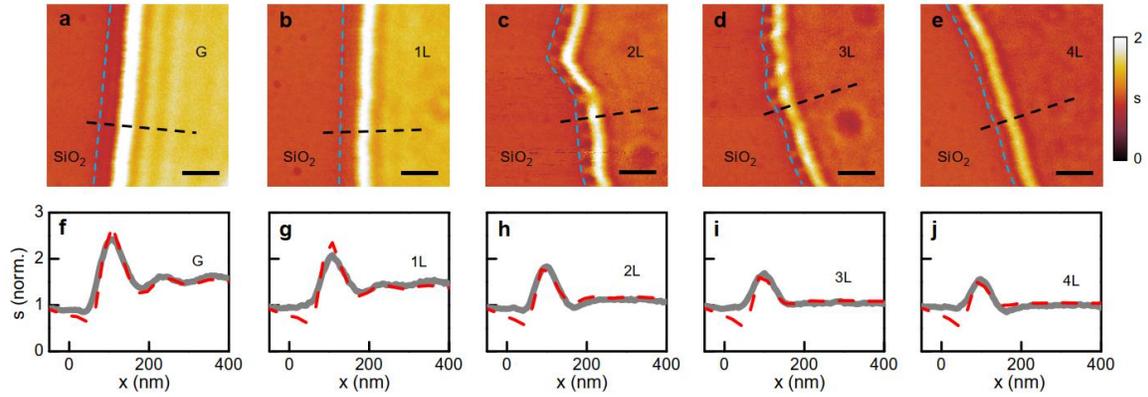

**Figure 2. a-e**, High-resolution IR amplitude images of bare graphene (G) and pentacene/graphene heterostructures (1L to 4L) with various pentacene thicknesses taken at $E = 116$ meV. The blue dashed curves mark the sample edges. The scale bars: 200 nm. **f-j**, The IR amplitude line profiles from both experiments (grey) and simulations (red). The experimental profiles were taken along the black dash lines in **a-e**. In all the images and profiles, the IR amplitude signal is normalized to that of the $SiO_2$/Si substrate.

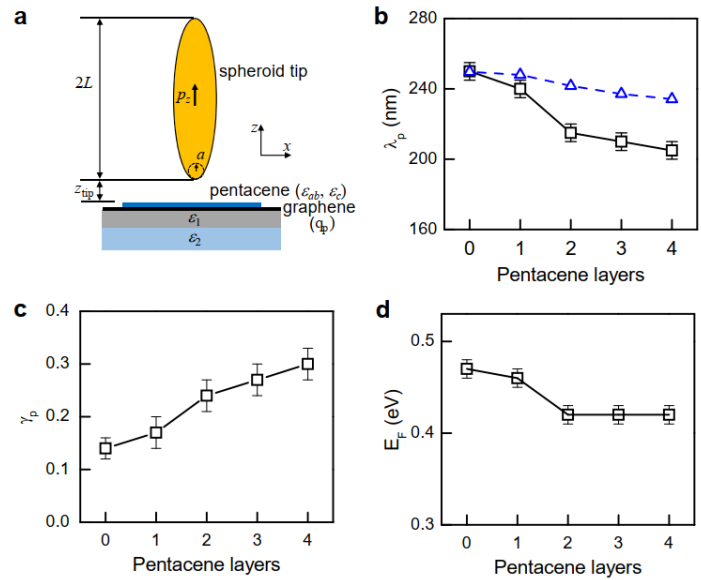

**Figure 3. a,** Illustration of the spheroid model with different parameters that we used to model the plasmon fringes profiles in Fig. 2f-j. **b,** The plasmon wavelength ($\lambda_p$) versus the number of pentacene layers from fringe profile fitting (black squares) and theoretical calculations (blue triangles) assuming a fixed Fermi energy of 0.47 eV. **c,** The plasmon damping rate ($\gamma_p$) versus the number of pentacene layers from fringe profile fitting. **d,** The Fermi energy of graphene versus the number of pentacene layers calculated based on the fitted $\lambda_p$ data (squares) in panel **b**. In panels **b-d**, the 0 pentacene layer corresponds to bare graphene.

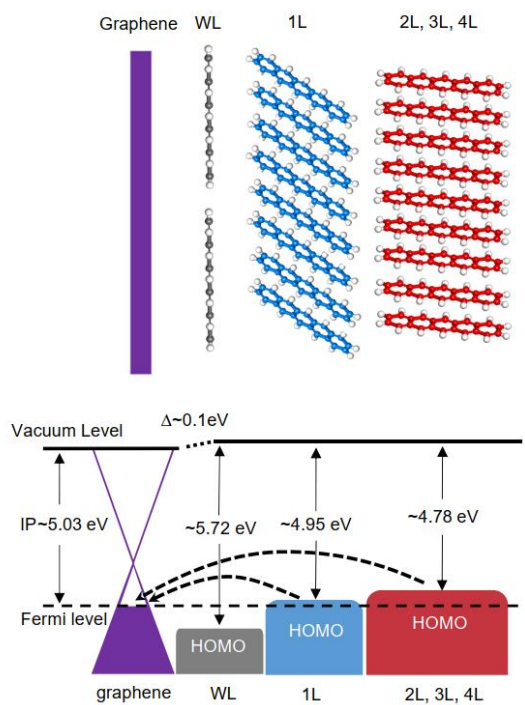

**Figure 4.** Illustration of the graphene/pentacene heterostructure (top) and the energy level alignment diagram (bottom). As thickness increases, pentacene changes from flat-laying wetting layer (WL), inclined one layer (1L) to standing few layers (2L, 3L, and 4L) relative to graphene, resulting in a lifting of pentacene HOMO energy level. The energy values listed in the alignment diagram are the ionization potentials (IP) of graphene and pentacene layers before electron transfer.

Supporting Information for

# Tailor plasmons in pentacene/graphene heterostructures with interlayer electron transfer

F. Hu[1,2]*, M. Kim[1,2]*, Y. Zhang[3], Y. Luan[1,2], K. M. Ho[1,2], Y. Shi[3], C. Z. Wang[1,2]†, X. Wang[3]†, Z. Fei[1,2]†

[1]Department of Physics and Astronomy, Iowa State University, Ames, Iowa 50011, USA
[2]U.S. DOE Ames Laboratory, Iowa State University, Ames, Iowa 50011, USA
[3]National Laboratory of Solid State Microstructures, School of Electronic Science and Engineering and Collaborative Innovation Center of Advanced Microstructures, Nanjing University, Nanjing 210093, China

* These authors contributed equally to this work.

† C.Z.W. (wangcz@ameslab.gov); X.W. (xrwang@nju.edu.cn); Z.F. (zfei@iastate.edu)

**List of contents**

1. Experimental details

2. Additional s-SNOM imaging data

3. Numerical modeling of the plasmon fringe profiles

4. Calculations of the plasmon wavelength

5. DFT calculation methods and results

6. Dielectric constants of pentacene

   Figures S1 – S9

1. **Experimental details**
To perform nano-infrared (IR) imaging studies of the pentacene/graphene heterostructures, we employed the scattering-type scanning near-field optical microscopy (s-SNOM). Our s-SNOM apparatus (Neaspec GmbH) is based on an atomic force microscope (AFM) operating in the tapping mode. Measurements were acquired at an AFM tapping frequency of $\Omega = 270$ kHz and a tapping amplitude of about 60 nm. As illustrated in Figure 1a, we utilized a metalized AFM probe, which is illuminated by a *p*-polarized mid-IR beam from a continuous-wave $CO_2$ laser (Access Laser). In our s-SNOM measurements, we used Arrow-NCPt probes from NanoAndMore. The radius of tip apex of these probes is about 25 nm that defines the spatial resolution of the s-SNOM. The standard observable of an s-SNOM experiment is complex scattering signal demodulated at the $n^{th}$ ($n = 3$ in the current work) harmonics of the AFM tip oscillation. We discuss mainly the amplitude part of the signal that is enough to describe the plasmonic responses of the samples.

Our pentacene/graphene vdW heterostructures were prepared by physical vapor transport deposition of uniform pentacene molecular layers on graphene. The heterostructure samples are sitting on the standard silicon wafers with 300-nm-thick thermal oxide on the top. The samples that we studied in this work include bare graphene, and one-layer (1L), two-layer (2L), three-layer (3L) and four-layer (4L) pentacene on graphene, determined by accurate AFM measurements (Figure S1).

2. **Additional s-SNOM imaging data**
In Figures S2 and S3, we present the excitation energy (*E*) dependent nano-IR amplitude images of graphene, 1L pentacene on graphene, and 2L pentacene on graphene. Here the IR amplitude is normalized to that of the $SiO_2$ substrate. From Figures S2 and S3, one can see that the IR contrast between the samples and the $SiO_2$ substrate shows a clear evolution with energy. This is mainly due to the increase of the substrate signal as *E* approaches the surface phonon resonance of $SiO_2$ at around 140 meV. Moreover, the fringe period or the fringe width of the samples shrinks with increasing laser energy, indicating smaller plasmon wavelength. This is consistent with the dispersion properties of graphene plasmons (Figure S5). In all laser energies, there is a small signal difference between bare graphene and 1L pentacene on graphene and a larger contrast between 1L and 2L pentacene on graphene. This is consistent with the results discussed in the main text (Figures 1-3).

Figure S4a,b present the nano-IR imaging data of two heterostructure samples at an excitation energy of $E$ = 110 meV. Sample 1 is the one that we extensively studied in the main text. Sample 2 is from a different wafer and it also contains bare graphene and 1L to 4L pentacene on graphene. Note that the excitation energy used here is slightly lower than that used in Figures 1 and 2 in the main text (116 meV). Here we compare the general signal contrast of different sample areas, which show good consistency among the two samples. For quantitative comparison, we plot in Figure S4c the average IR amplitude at the sample interior versus the number of pentacene layers (0 layer corresponds to bare graphene). Here one can see that the general trend of the signal evolution with layer thickness is consistent in the two heterostructure samples. There is a slight difference in 1L pentacene on graphene, which is possibly due to the degradation of 1L pentacene on sample 2 that leads to nonuniform signal distributions (Figure S4b).

## 3. Numerical modeling of the plasmon fringe profiles

To model the fringes profiles of plasmons confined inside graphene or pentacene/graphene heterostructures, we model our AFM tip as an elongated metallic spheroid (see Figure 3a in the main text): the length of the spheroid is 2$L$ and the radius of curvature at the tip ends is $a$. Here, $a$ is set to be 25 nm according to the manufacturer and $L$ is set to be 500 nm and it is not a very sensitive parameter so long as $L \gg a$. The scattering amplitude $s$ (before demodulation) scales with the total radiating dipole $p_z$ of the spheroid. Therefore, to fit the line profiles perpendicular to the fringes inside samples, we need to calculate $p_z$ at different spatial coordinates ($x$, $z$) of the lower end of the AFM tip. Here, $x$ is the in-plane coordinate perpendicular to samples and $z$ is the out-of-plane coordinate perpendicular to the sample surface. By calculating $p_z$ at different $z$, we can perform 'demodulation' of the scattering amplitude $s$ and get different harmonics of the scattering signal and calculating $p_z$ at different $x$ allows us to plot the modeling profiles of IR amplitude. In all our simulations, we assume no position dependence in the $y$-direction for simplicity. The dielectric constants of SiO$_2$ used in the calculations are adopted from literature.[1] The dielectric constants of pentacene layers ($\varepsilon_{ab}$, $\varepsilon_c$) used in the calculations given in Section 6 below. The key modeling parameters for graphene are the plasmon wavelength ($\lambda_p$) and damping rate ($\gamma_p$). By fitting the experimental plasmon fringe profiles (Figure 2f-j in the main text), we can determine accurately $\lambda_p$ and $\gamma_p$ based on the experimental data (Figure 3b,c in the main text).

## 4. Calculations of the plasmon wavelength

In order to determine the Fermi energy ($E_F$) of graphene, we need to calculate $\lambda_p$ theoretically and then compared to the experimental result obtained from fringe profile fitting (see the section above). For that purpose, we first compute the plasmon dispersion colormaps (Figure S5) by evaluating numerically the imaginary part of the reflection coefficient Im($r_p$) for the entire pentacene/graphene/substrate heterostructure system by using the transfer matrix method. These colormaps reveal the photonic density of states (DOS), and the plasmonic mode appears as a bright curve revealed by the colormaps. Such a dispersion calculation method has been widely applied in the studies of graphene plasmons and other types of polaritons. The optical conductivity of graphene is obtained by the Radom phase approximation methods (see Ref. 11 in the main text). The dielectric constants of pentacene layers used in the calculations given in Section 6 below. The dielectric constants of $SiO_2$ used in the calculations are adopted from literature.[1] Based on the calculated dispersion colormaps, we can determine the plasmon wavevector $q_p$ and hence the plasmon wavelength $\lambda_p = 2\pi/q_p$ at any excitation energy. The dispersion colormaps shown in Figure S5 are calculated with different choices of $E_F$ values shown in Figure 3d in the main text. The $\lambda_p$ values read out from these colormaps match well the $\lambda_p$ results determined through fringe profile fitting (Figure 2 in the main text), which confirms the validity of $E_F$ values shown in Figure 3d.

## 5. DFT calculation methods and results

We performed first-principles electronic structure calculations based on density functional theory (DFT) using Vienna *ab initio* simulation package (VASP).[2,3] We employed the projector augmented wave method[4] and a plane-wave basis set with 400 eV energy cutoff. For the exchange-correlation functional, we used Perdew-Burke-Ernzerhof (PBE) functional[5], and a total of 5×2×1 *k* point meshes were used. We included the van der Waals energy using the DFT-D3 method.[6] In our slab calculations for the pentacene thin films, we used sufficiently thick vacuum regions (> 23 Å) to prevent the unwanted interactions between periodic images. We considered the structural model of pentacene layers that consist flat-lying wetting layer (WL) on graphene, the inclined one layer (1L) and the standing two (2L) and three layers (3L) (Figure S6).[7] The dielectric matrix was determined using density functional perturbation theory. In Figures S7 and S8, we presented the calculated band structures of pentacene layers and their potential energy line profiles taken along the vertical direction. From these calculations, we can determine the ionization potential (IP) of all pentacene layers (Table 1 in the main text).

Due to the self-interaction error, the conventional DFT calculations with local density approximation (LDA) or generalized gradient approximation (GGA) are not supposed to give accurate valence band energies, which in turn results in underestimated IP values.[8] Therefore, we performed the GW calculation (in the $GW_0$ level) that is expected to give a reasonable value for the IP. Due to the large computational cost, we used the approximation where we estimate the shift of the valence band maximum by that of the highest valence band energy at the Γ point of the Brillouin zone ($\Delta_{VB,\Gamma}$) in the 3D bulk pentacene crystal (Figure S9a ). In the GW calculation of the bulk pentacene, we used 2×2×1 *k* point meshes. We considered further correction coming from limited number of empty bands ($\Delta_{eb}$). Specifically, we estimated the valence band energy with the infinite number of the empty bands ($N_{eb} = \infty$) included in the calculation by fitting the results of a series of different $N_{eb}$ to the formula $A/N_{eb} + B$.

The results of both corrections are shown in Figure S9b,c, respectively. We denoted the sum of the two correction terms by $\Delta_{GW}$ (i.e., $\Delta_{GW} = \Delta_{VB,\Gamma} + \Delta_{eb}$). The total GW correction is as large as ≅ 0.96 eV. After corrections, the IPs of thin pentacene layers are calculated to be 5.72 eV, 4.95 eV, 4.78 eV, and 4.78 eV for WL, 1L, 2L and 3L pentacene, respectively (Table 1 in the main text). These results are consistent with previous experiments.[9] Based on the calculation results, we conclude that the IP of pentacene is solely dependent on the orientation angle of pentacene molecules. Therefore, we expect that IPs of thicker pentacene layers (4L or above) are all about 4.78 eV.

## 6. Dielectric constants of pentacene

The dielectric constants of pentacene layers we used for the fringe profile modeling and dispersion calculations are also from DFT calculations. With DFT, we calculated the static dielectric constants of pentacene layers with different thicknesses. It is known from the previous study[10] that pentacene has a flat dielectric response up to the visible region when it is away from the strong vibrational modes of pentacene (the nearest strong resonance is at about 112 meV with a resonance width of 0.4 meV). Therefore, it is appropriate to use the static dielectric constants for calculations in the mid-infrared region. The calculated in-plane dielectric constants ($\varepsilon_{ab}$) for 1L, 2L and 3L pentacene are about 2.1, 2.6 and 2.7, respectively. The out-of-plane dielectric constants ($\varepsilon_c$) for 1L, 2L and 3L pentacene are about 1.3, 2.0 and 2.6, respectively. Both $\varepsilon_{ab}$ and $\varepsilon_c$ increase with pentacene thickness and they are trending towards the value for bulk pentacene films: $\varepsilon_{bulk} \approx 3.0$.[10] For 4L pentacene, we used dielectric constants of 3L pentacene as an approximation. The calculated $\lambda_p$ only varies a little (~1.5%) even using $\varepsilon_{bulk} \approx 3.0$ for 4L pentacene. There is in fact a small *ab*-plane anisotropy (about 1%, 5% and 12% for 1L, 2L and 3L pentacene, respectively) in the dielectric constants according to our calculations, but it only causes tiny variations to $\lambda_p$ (about 0.05%, 0.3% and 0.8% for 1L, 2L and 3L pentacene, respectively) due to the nanoscale thicknesses of the pentacene layers. Therefore, we used averaged *ab*-plane dielectric constants in our calculations ($\varepsilon_{aa} + \varepsilon_{bb}$)/2.

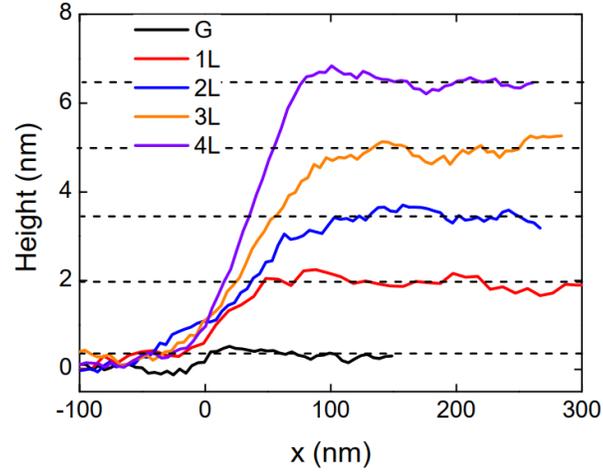

**Figure S1.** The AFM topography profiles of graphene (G) and 1L to 4L pentacene on graphene extracted from Figure 1b in the main text. Here the 0 pentacene layer corresponds to bare graphene.

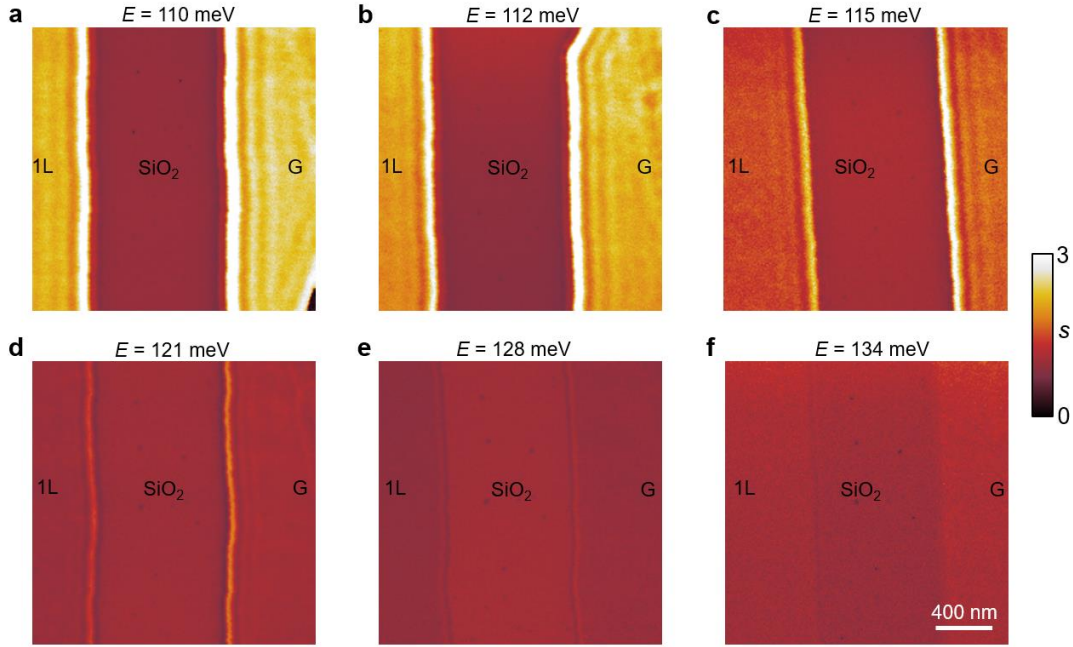

**Figure S2.** Excitation laser energy dependent nano-IR imaging data of bare graphene (G) and 1L pentacene on graphene. Here we plot the IR amplitude normalized to that of the $SiO_2$ substrate.

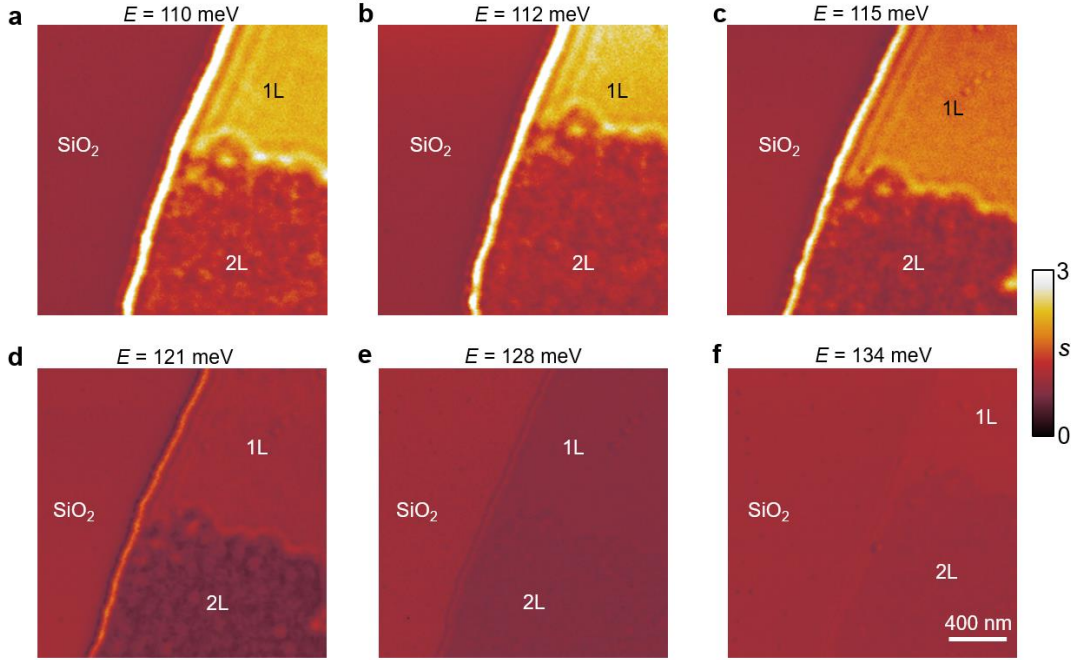

**Figure S3.** Excitation laser energy dependent nano-IR imaging data of 1L and 2L pentacene on graphene. Here we plot the IR amplitude normalized to that of the SiO$_2$ substrate.

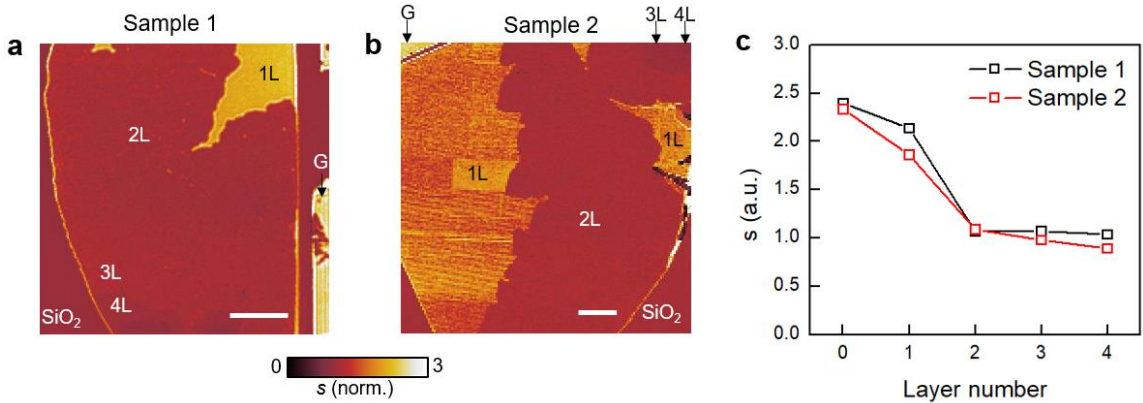

**Figure S4. a,b,** Nano-IR imaging of two samples at an excitation energy of 110 meV (slightly lower energy compared to that used in Figure 1 and 2 in the main text). Sample 1 is the sample we studied extensively in the main text. Sample 2 is a different sample on a different wafer. Scale bars: 2 μm. **c,** The IR amplitude signals of the two samples taken from **a,b** versus the number of pentacene layers.

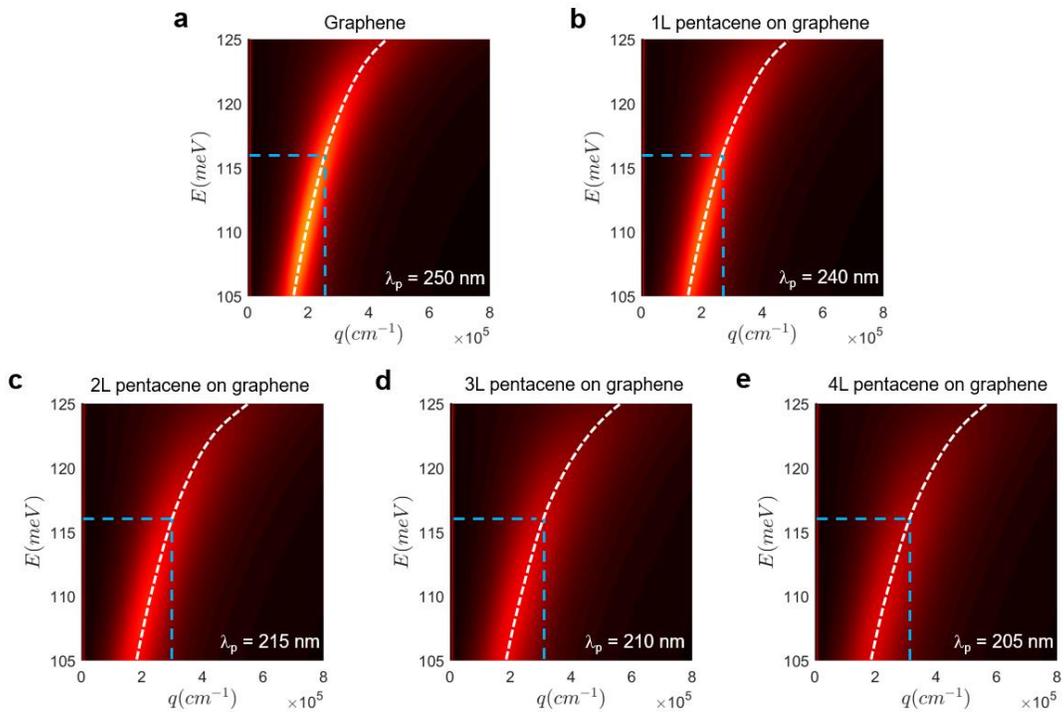

**Figure S5.** Calculated dispersion colormaps of bare graphene (**a**) and pentacene layers with different thicknesses on graphene (**b**-**e**). The white dashed curves mark the dispersion relation of graphene plasmons revealed by the color maps. The horizontal and vertical blue dashed lines mark the excitation energy ($E$ = 116 meV) and corresponding plasmon wavevector determined by the dispersion diagrams.

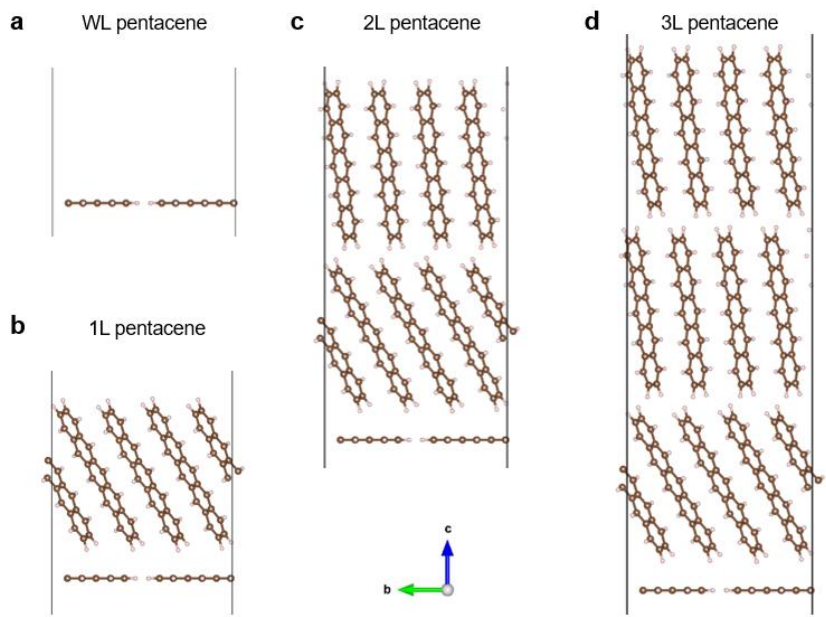

**Figure S6.** Atomic structures of different pentacene layers that we constructed for DFT calculations.

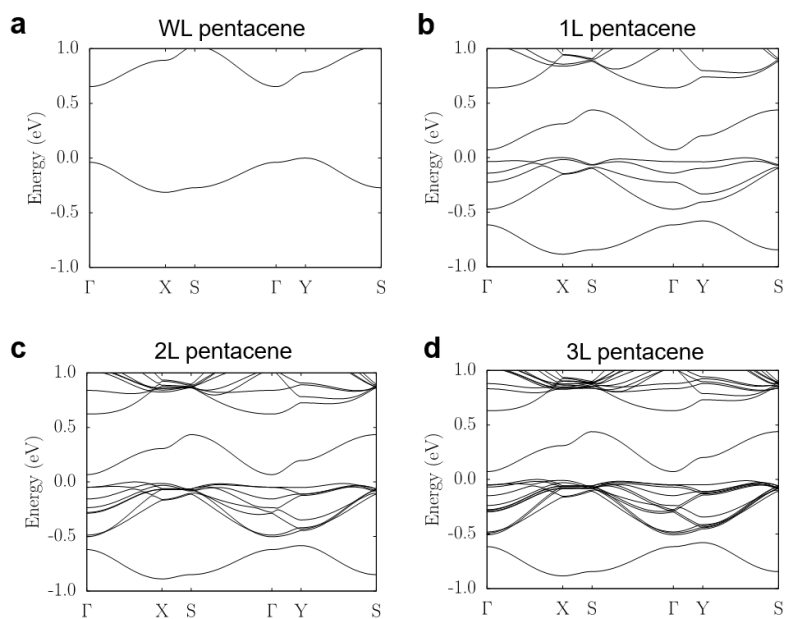

**Figure S7**. The DFT calculations of band structures of WL, 1L, 2L and 3L pentacene. In all the plots, the band structures are shifted on purpose to set the valence band maximum right at 0 eV.

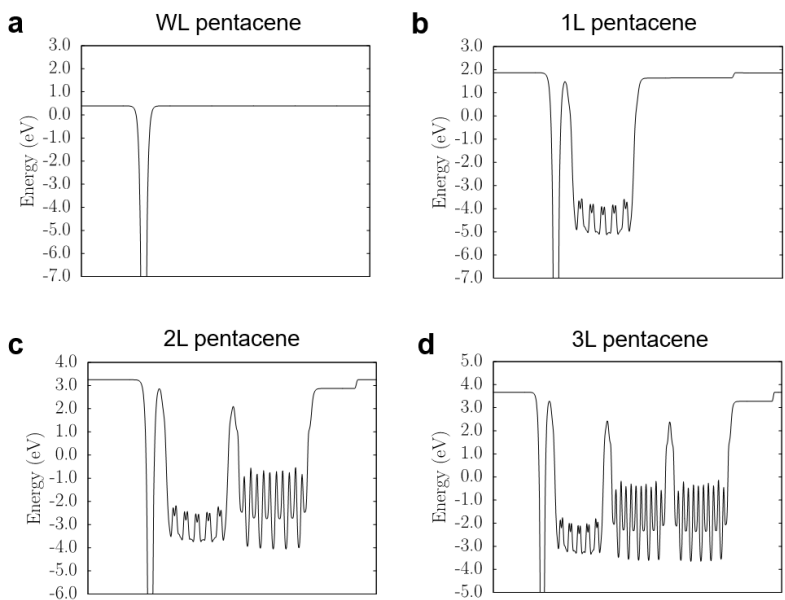

**Figure S8.** The DFT calculations of potential energy profiles of WL, 1L, 2L, and 3L pentacene along the $c$ axis (perpendicular to the pentacene layers).

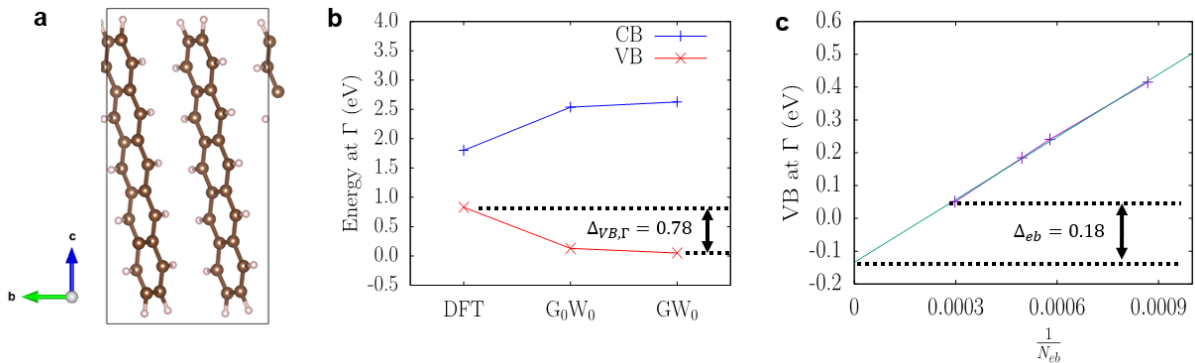

**Figure S9**. **a**, Unit cell of bulk pentacene that we used for GW calculations. **b**, Shift of the valence band maximum by that of the highest valence band energy at the $\Gamma$ point of the Brillouin zone ($\Delta_{VB,\Gamma}$). **c**, Correction coming from limited number of empty bands ($\Delta_{eb}$).